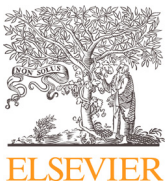
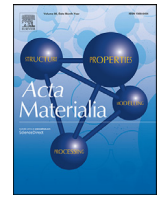

# The influence of structural variations on the constitutive response and strain variations in thin fibrous materials

Mossab Alzweighi[a], Rami Mansour[a], Jussi Lahti[b,c], Ulrich Hirn[b,c], Artem Kulachenko[a,c,*]

[a] *Solid Mechanics, Department of Engineering Mechanics, KTH Royal Institute of Technology, SE-100 44 Stockholm, Sweden*
[b] *Institute of Bioproducts and Paper Technology, Graz University of Technology, Inffeldgasse 23, 8010 Graz, Austria*
[c] *CD Laboratory for Fiber Swelling and Paper Performance, 8010 Graz, Austria*

**a b s t r a c t**

The stochastic variations in the structural properties of thin fiber networks govern to a great extent their mechanical performance. To assess the influence of local structural variability on the local strain and mechanical response of the network, we propose a multiscale approach combining modeling, numerical simulation and experimental measurements. Based on micro-mechanical fiber network simulations, a continuum model describing the response at the mesoscale level is first developed. Experimentally measured spatial fields of thickness, density, fiber orientation and anisotropy are thereafter used as input to a macroscale finite-element model. The latter is used to simulate the impact of spatial variability of each of the studied structural properties. In addition, this work brings novelty by including the influence of the drying condition during the production process on the fiber properties. The proposed approach is experimentally validated by comparison to measured strain fields and uniaxial responses. The results suggest that the spatial variability in density presents the highest impact on the local strain field followed by thickness and fiber orientation. Meanwhile, for the mechanical response, the fiber orientation angle with respect to the drying restraints is the key influencer and its contribution to the anisotropy of the mechanical properties is greater than the contribution of the fiber anisotropy developed during the fiber sheet-making.

## 1. Introduction

Bio-based fibrous materials are broadly used in a number of industrial applications [1] of which paper and board-making are the largest by volume and total revenue. The usage of these materials has gradually advanced from conventional to more sophisticated applications where they are used as substrates for sensors, actuators and circuit boards [2]. The biodegradability, renewability, efficient production technique and low prices provide the preference of using those materials over synthetic materials like glass and plastic [3]. Furthermore, a key advantage of these materials is the ability to control the anisotropic properties and enhance the performance of the material in the loading direction [4].

Paper-based materials, for which fibers are the main constitutive components, are characterized by randomness in their mechanical response. This randomness is originated in the production process, in which fibers, fiber segments, and other chemical substances are combined into a thick solution [5]. The solution is thereafter impinged into a wired network where it is compressed and drained continuously. Due to this process, the produced sheet of paper is inherently heterogeneous [6,7] and presents variations in the structural properties [8].

The stochastic heterogeneity of paper materials and its characterization have been subjected to extensive studies. In particular, density [9], thickness [10], anisotropy [11] and fiber orientation [12] vary locally in the machine-made materials. It is noted that these structural properties may be spatially correlated. For instance, a correlation between density and thickness has been observed [13]. Although it has been demonstrated that paper specimens with higher average grammage present higher strength and stiffness [14], the effect of local spatial variations on the global mechanical response is less understood. The local inhomogeneity in fiber network topography has been shown to be the main reason for the non-affine deformation [15] in paper-based materials and has a key role in the failure mechanism of the network [16]. Local regions with higher grammage present higher strength and stiffness [17] while lower grammage show higher local tensile strains that are correlated to the locus of failure initiation [18]. Moreover, in an experimental investigation, the influence of local structural properties on the local strain field has been quantified, showing that the thickness, basis weight, density, and local fiber orientation combined can justify 39% of the total variation in the local strain [19].

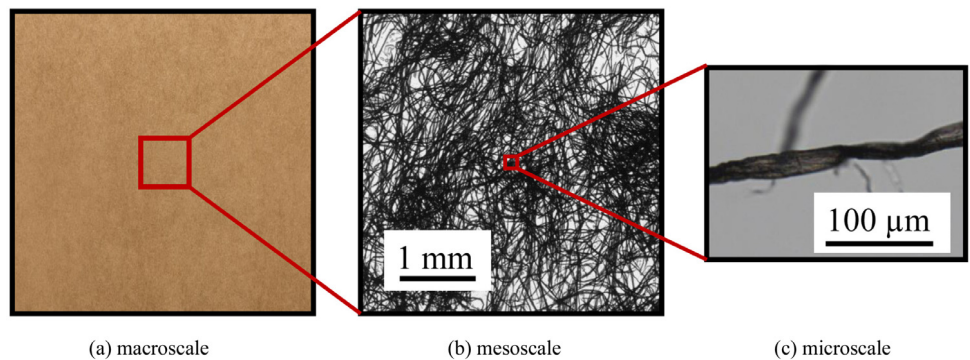

(a) macroscale　　　　(b) mesoscale　　　　(c) microscale

**Fig. 1.** Multiscale of the fibrous materials.

Using full 3D fiber-network generation and simulations, the strain localization pattern was found to vary randomly as does the global mechanical response [20]. The complex inhomogeneous nature of paper materials requires the study of different scales in order to model the material behavior [21]. Three discrete scales can be recognized: microscale (fiber scale), mesoscale (network scale), and macroscale (sheet scale) [21]. The fiber scale tackles the mechanical response of a single fiber and its geometrical properties (see Section 3.1). The mesoscale is the size of the fiber network for which a continuum-based model with effective properties is proposed. In this study, it is defined by a network size of 4 × 4 mm$^2$ (see Section 3.2). The three different scales for the studied paper-based material is shown in Fig. 1.

To bridge the gap between the scales of heterogeneous materials, multiscale and homogenization approaches are widely used for composites [22–25] and were recently reviewed in [26]. The heterogeneous fiber networks received less attention with some examples with collagen networks [27] and paper [28]. In [29], the authors employed a stochastic multiscale approach consisting of microscale, mesoscale, and macroscale to model an isotropic fibrous sheet to capture the strain localization. The effect of drying was not accounted for. The data for the spatial strain and strength to failure fields were taken from the artificially created sheets rather than from experiments. Those fields have been shown to be the main predictors of the random strain localization pattern.

The aforementioned studies lack the ability to identify the individual impact of the structural properties on the local strain field as well as on the global mechanical response. In experimental studies, singling out the influence of one structural property by imposing the other properties to be constant is elusive. Also, using micromechanical models to investigate the heterogeneity in the material structure is computationally formidable at the product scale. Overcoming these limitations is crucial for proper reliability analysis [30,31] and product scale design [32] of fibrous materials considering local random variability.

In this work, the impact of spatial structural variability is assessed using a combined experimental, numerical and continuum model approach. The variability is first quantified through experimental local measurements of the thickness, density, fiber orientation and fiber anisotropy. Micromechanical fiber network simulations, with fiber properties derived from experimental observations, are thereafter employed to study the influence of the structural properties on the mesoscale mechanical response. A novelty in the micromechanical simulations is the ability to account for the impact of the restrained drying condition during the production process on the fiber properties. A continuum model of the mesoscale mechanical response is thereafter proposed with material parameters that are function of the local properties. The impact of variability is finally assessed by applying the measured spatial fields of thickness, density, fiber orientation and fiber anisotropy in a macromechanical Finite-Element (FE)-model where each element is governed by the proposed mesoscale continuum model. The accuracy of the proposed approach with regard to the prediction of the mechanical response and spatial strain field under uniaxial loading is evaluated by comparison to experimental measurements.

## 2. Experimental

### 2.1. Measurement of local strain

In this experimental study, four sack paper specimens were exposed to a uniaxial tensile load in the Machine Direction (MD) with a strain rate of 40 mm/min under a standard climate of 23°°C and 50% relative humidity. Their local strain variation is recorded just prior to uniaxial tensile rupture which is estimated at 2.2% strain. The full size of the specimens is 40 × 65 mm$^2$ with a Region of Interest (RoI) of 32 × 56 mm$^2$ defined inside the deformed samples as seen in Fig. 2. The local strain was measured with Digital Image Correlation (DIC) [33] with a resolution of 0.74 mm/pixel. The local strain map within RoI was resized and aligned with the measured structural property maps described in Section 2.2 with a resolution of 1 mm/pixel by utilizing a landmark based registration and a shape preserving coordinate transform [34].

### 2.2. Measurement of thickness and density

After determining the local strain, the local thickness was measured with a twin laser profilometer [35] with a resolution of 100 μm/pixel within RoI. Local basis weight was determined with a $\beta$-radiographic transmission method [36] with a resolution of 50 μm/pixel within RoI. The measured thickness and basis weight maps were then aligned with each other with a resolution of 1 mm/pixel. Finally, the density map was calculated by dividing the aligned basis weight map with the aligned thickness map.

### 2.3. Measurement of fiber orientation and anisotropy

The local fiber orientation was measured with a sheet splitting method [37]. Each specimen was split in the thickness direction into 14–20 layers and subsequently scanning the layers by an optical scanner with a resolution of 13 μm/pixel. For each layer, the Fiber Orientation Distribution (FOD) is defined for subsets of 1 × 1 mm$^2$. The fiber orientation map for all the layers in the local 1 × 1 mm² subsets was then determined by averaging the FOD across the layers in the thickness direction and subsequently fitting ellipses to the averaged distributions.

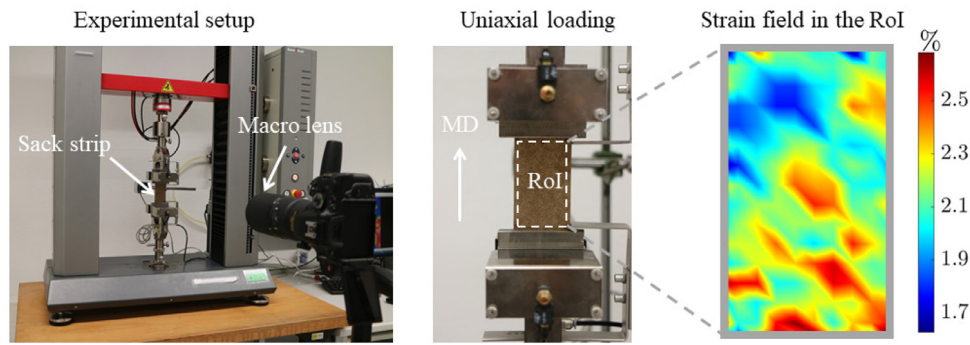

**Fig. 2.** Principle of measuring local strain of the studied specimens in MD under uniaxial tensile loading.

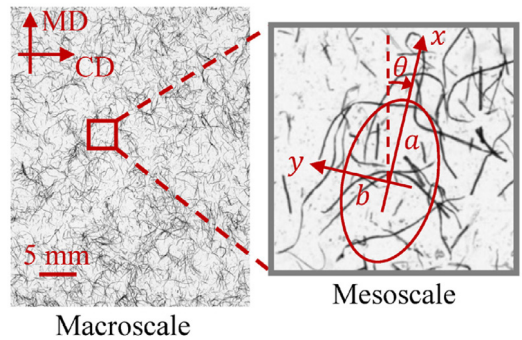

**Fig. 3.** Principle of determination of local fiber orientation angle $\theta$ and anisotropy $\lambda$ for one sheet.

The local Fiber Orientation Distribution (FOD) is defined by two parameters, orientation and anisotropy, derived from the direction and shape of the fitted ellipse as described in Section 3.2, see Fig. 3. The orientation angle $\theta$ is defined as the deviation of the local major orientation direction from the MD. The anisotropy is computed as

$$\lambda = 1 - b/a, \quad (1)$$

where $a$ and $b$ are the major and minor axes of the local fiber orientation distribution ellipse, respectively.

The local variations in the measured structural properties as well as the strain fields are shown in Fig. 4 for two of the studied specimens. In the remainder of this paper, a methodology for quantifying the individual impact of local variability of each of the studied structural properties on both the strain field and mechanical response is presented.

## 3. Fiber network simulation

The sampling of the material data on the mesoscale from the experiments has fundamental difficulties such as a laborious way of analyzing the data by splitting, the need for testing in different directions beyond the yield stress, and performing well-controlled tests on relatively small samples. In order to avoid these limitations, we used a micromechanical simulation framework. In this framework, we reconstruct the fiber network with a controlled density, thickness, and fiber anisotropy. This enables performing the mesoscale (network scale) mechanical analyses in various directions, varying degrees of anisotropy, and yet maintaining other structural properties constant.

### 3.1. Microscale modeling

In order to study the influence of structural variability, we have employed micromechanical fiber network simulations. The geometries of the fiber including fiber length, width, shape factor, wall thickness and width to high ratio, are acquired using a characterization methodology outlined in [4]. These geometrical data are extracted based on measurements of wet kraft pulp using Fiber Morphology Analysis (FMA) [38]. The cross-sectional data are corrected by processing the scanned images of the kraft sheet with the Micro-Computed tomography ($\mu$CT) [39]. The pulp used to produce the sheets was an unbleached (Kappa value is 43) softwood Kraft chemical pulp fibers being a mixture of spruce (~80%) and pine (~20%) wood and was characterized with the methods described in [4]. The relevant pulp data are listed in Table 1.

The network generation is performed by randomly depositing the fibers [40] in a planar surface where the curvature of the fiber is constant and parallel to the deposition plane. During the deposition, the intersection between the fibers is determined and the intersection point is lifted up to avoid penetration, see Fig. 5(a). The fiber geometry is thereafter smoothed in order to avoid kinking. This random deposition of fibers is continued until the specified grammage is reached. The density of the softwood pulp reported in the literature ranges between 1200 kg/m³ [41] to 1500 kg/m³ [42] and is dependent on the lignin content. The adopted value used during the deposition in this study is 1400 kg/m³. The code for generating the network along with the documentation is available as supplementary material in [29].

The fibers are represented using a finite-element model by a series of 3-node 3D Timoshenko beam elements. The cross-section is circular with a solid or hollow cross-section. The pointwise contact beam to beam with traction and separation law is used to model the bond between fibers [40]. In this work, we did not consider debonding between the fibers as this phenomenon does not significantly influence the stress-strain curves in the considered strain interval before the ultimate failure and softening [43]. The softening behavior was not included in the constitutive model either. Not including them can, however, prohibit capturing accurately the strain field locally as the strain localization and the associated bond failures can occur prior to the global failure. The constitutive response of the fiber is modeled using bilinear plasticity with an isotropic hardening law. The flow stress is given by $\sigma_{s,f} + E_{tan,f}\, \varepsilon_{e,pl}$ where $\sigma_{s,f}$ is the initial yield stress of the fiber, $E_{tan,f}$ is the tangent modulus and $\varepsilon_{e,pl}$ is the equivalent plastic strain. The latter is incrementally computed from the plastic stain increments $\boldsymbol{d\varepsilon}_{pl}$ as $d\varepsilon_{e,pl} = (\frac{2}{3}\boldsymbol{d\varepsilon}_{pl}^T \boldsymbol{d\varepsilon}_{pl})^{1/2}$ which has three components (one normal and two transverse shear strains) in the used Timoshenko beam element. The material properties of the fibers are determined in Section 3.3.

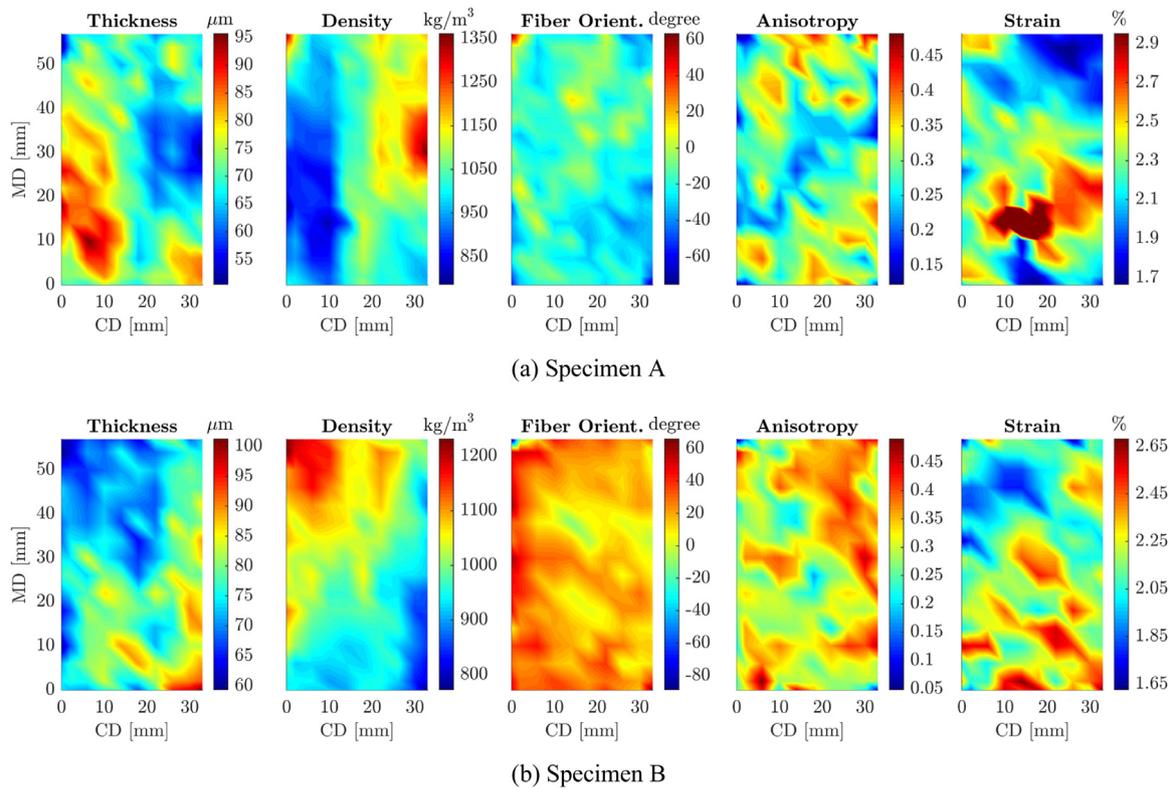

**Fig. 4.** Measured local thickness, density, fiber orientation angle and fiber orientation anisotropy as well as local strain field for two specimens under uniaxial loading. (For interpretation of the references to colour in this figure legend, the reader is referred to the web version of this article.)

**Table 1**
The geometrical properties of fibers used in the micromechanical simulation tool.

|  | Mean value | Standard deviation | Source of data |
| --- | --- | --- | --- |
| Fiber width [μm] | 17.38 | 7.27 | FMA |
| Width to height ratio [-] | 2.9 | 1.72 | μCT |
| Fiber length [mm] | 2.33 | 1.34 | FMA |
| Fiber shape factor [-] | 0.945 | 0.015 | FMA |
| Wall thickness [μm] | 5.28 | 2.18 | μCT |

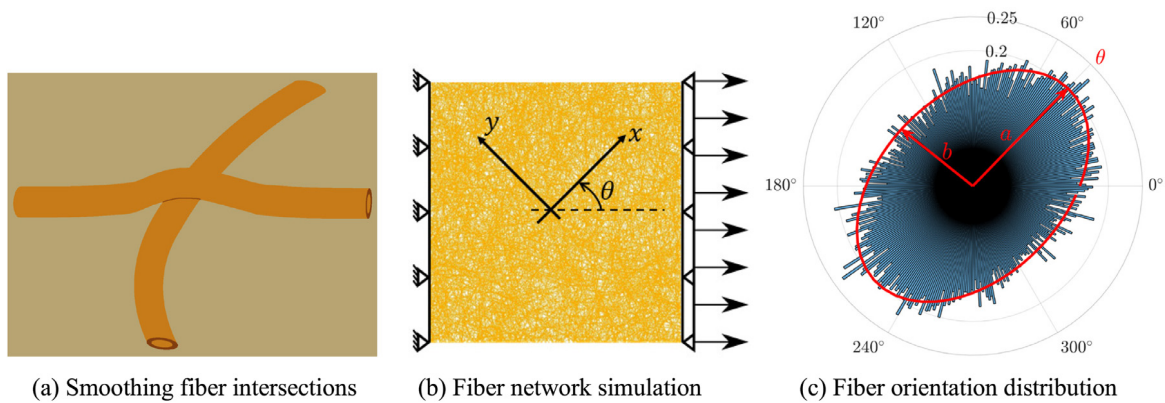

(a) Smoothing fiber intersections    (b) Fiber network simulation    (c) Fiber orientation distribution

**Fig. 5.** A $4 \times 4$ mm$^2$ fiber network model with a target degree of anisotropy $\lambda = 1 - b/a$.

### 3.2. Mesoscale simulation

Direct simulations of $4 \times 4$ mm$^2$ fiber networks with grammage 70 g/m$^2$ are used to quantify the influence of fiber orientation and degree of anisotropy on the constitutive response. The chosen $4 \times 4$ mm$^2$ size is small enough to be representative of the mesoscale, i.e. to represent a discretization size that captures local variation in the macroscale. The relatively large size of the mesoscale ensures that 1) random realizations having the same thickness, density, fiber orientation and degree of anisotropy result in approximately the same constitutive response and 2) the influence of the boundary condition on the uniaxial response of the mesoscale fiber network is small.

The effect of fiber orientation $\theta$ on the constitutive response is studied using three uniaxial tests in $\theta = 0$, $\theta = 45°$ and $\theta = 90°$ with respect to the loading direction as seen in Fig. 5(b). These tests are performed for different degrees of anisotropy ranging from $\lambda = 0.15$ to $\lambda = 0.7$, while maintaining the same average density of the element. The anisotropy is altered by controlling the orientation of the constitutive fibers during the deposition phase (refer to Section 3.1). It is defined based on the polar plot of the Fiber Orientation Distribution (FOD), see Fig. 5(c), which is computed from the fibers that constitute the generated network. An elliptic equation in polar coordinates is fitted to the FOD [44] with respect to the major and minor axis of the ellipse, $a$ and $b$, respectively, and the fiber orientation $\theta$ as seen in Fig. 5(c). The anisotropy is thereafter computed from Eq. (1).

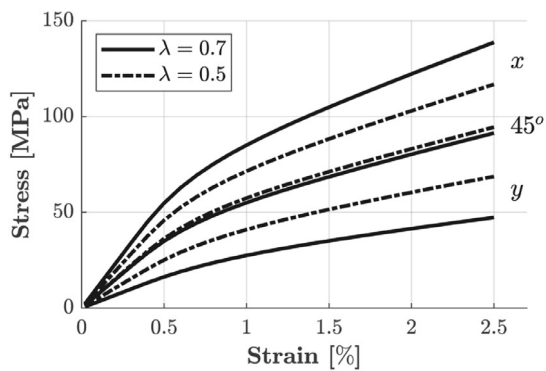

(a) Anisotropic fiber networks

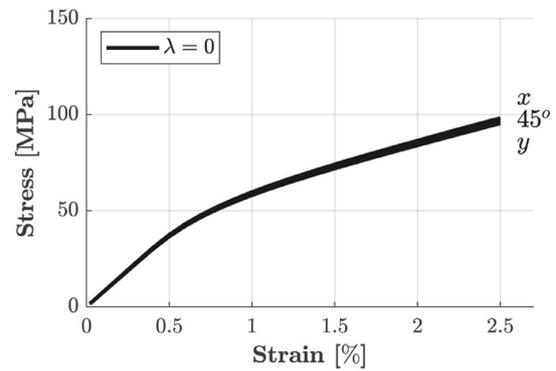

(b) Isotropic fiber network

**Fig. 6.** Mesoscale stress-strain curves from direct simulation for different fiber orientations and degrees of anisotropy.

The response of the uniaxial tests in Fig. 5(b) for two different degrees of anisotropy, is shown in Fig. 6(a). As can be seen, the difference in response between the $x$ and $y$ directions expectedly decreases with decreased degree of anisotropy, while the 45° direction remains almost unchanged. For the isotropic case, see Fig. 6(b), a similar response is observed in all the three directions. These results do not take into account the effect of restrained drying in MD, which is modeled in Section 3.3.

### 3.3. Effect of restrained drying condition during the production process

During the papermaking process, the fiber web, which is mainly constituted of fibers, is drained and the fiber network consolidation takes place. In the sack paper investigated in this study, the shrinkage is restrained in MD while free shrinkage is allowed in Cross Direction (CD) during the drying. This drying process is well known to influence the properties of the produced materials [45,46], by increasing the stiffness [47] in the restrained drying direction and reducing it in the free shrinkage direction. The physics behind this change is attributed to the impact on the microstructure of the fibers with micro-compressions and realignment of the microfibrils inside the fiber [47,48]. Consequently, in a sheet subject to drying, the fiber properties will change according to their orientation with respect to the drying direction [49].

In the direct simulation tool, the effect of restrained drying in MD and free shrinkage in CD is implemented by modeling the fiber property as a function of its orientation angle $\theta_f$ with respect to MD according to

$$E_f(\theta_f) = E_f^{MD}\cos^4\theta_f + E_f^{CD}(1 - \cos^4\theta_f), \qquad (2)$$

where $E_f$ is the fiber elastic modulus, $E_f^{MD}$ and $E_f^{CD}$ are fitting parameters representing the elastic modulus of a fiber oriented in MD and CD, respectively, see Fig. 7. The selection of this function, although being empirical, is motivated by the analytical transformations presented in ref. [50]. The same relation is assumed to describe the fiber shear modulus $G_{xy,f}(\theta_f)$, yield strength $\sigma_{s,f}(\theta_f)$ and tangent modulus $E_{tan,f}(\theta_f)$ with corresponding fitting parameters listed in Table 2. These are found by matching the fiber network responses to the experimental uniaxial responses in MD, CD and 45° directions.

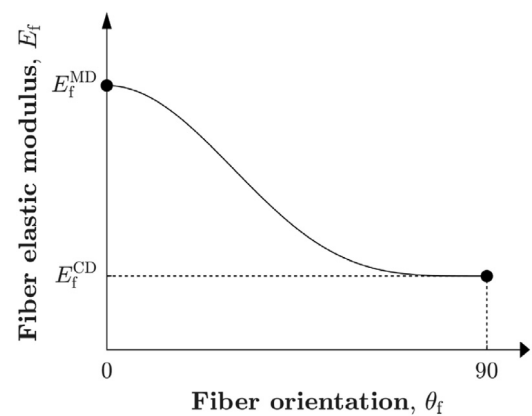

**Fig. 7.** Fiber elastic stiffness $E_f$ as a function of the fiber orientation angle.

It is noted that Eq. (2) and the corresponding fitting parameters from Table 2 are only valid for the case of restraint drying in the MD and freely dried in the CD. This is typical for the commercially produced paper web with the exception of the web edges, where the effect of constraints may affect the shrinkage in the CD [51].

From the experimentally studied specimens, it was found that the average anisotropy is $\lambda=0.34$. Therefore, a fiber network with the same degree of anisotropy ($\lambda=0.34$) was generated using the micromechanical tool. The mechanical response of this fiber network was fitted to the stress-strain curves for the same material of the sack paper. Once the MD response was fitted, the CD and 45° responses were computed. Fig. 8(a) shows that with the measured anisotropy alone we were unable to achieve a reasonable agreement with the experiment conducted in all the directions. However, by adding the influence of drying a good fit was achieved, see Fig. 8(b). This is explained as follows. As Fig. 7 shows, the selected function for the angular dependency of the fiber properties is not symmetric with respect to 45° angle and is flatter toward the CD. This means the properties of the fibers change faster toward those in the CD in order to capture the experimental stress-strain curves in all directions as in Fig. 8(b). Neglecting this angular dependency with respect to the drying direction and assuming the same fiber properties regardless of orientation results in the mismatch against the experiment shown in Fig. 8(a). This result shows the impor-tance of including the effect of drying as well as the impact it has compared to the anisotropy alone. It is concluded that the effect of the drying has a greater contribution to the anisotropy in the mechanical properties of the sheet compared to the effect of the fiber anisotropy alone.

Table 2
Experimentally determined model parameters.

| Elastic modulus [GPa] | | Shear modulus [GPa] | | Yield stress [MPa] | | Tangent modulus [GPa] | |
|---|---|---|---|---|---|---|---|
| $E_f^{MD}$ | $E_f^{CD}$ | $G_{xy}^{MD}$ | $G_{xy}^{CD}$ | $\sigma_{s,f}^{MD}$ | $\sigma_{s,f}^{CD}$ | $E_{tan,f}^{MD}$ | $E_{tan,f}^{CD}$ |
| 45 | 18.3 | 9 | 3.7 | 240 | 98 | 11 | 0.9 |

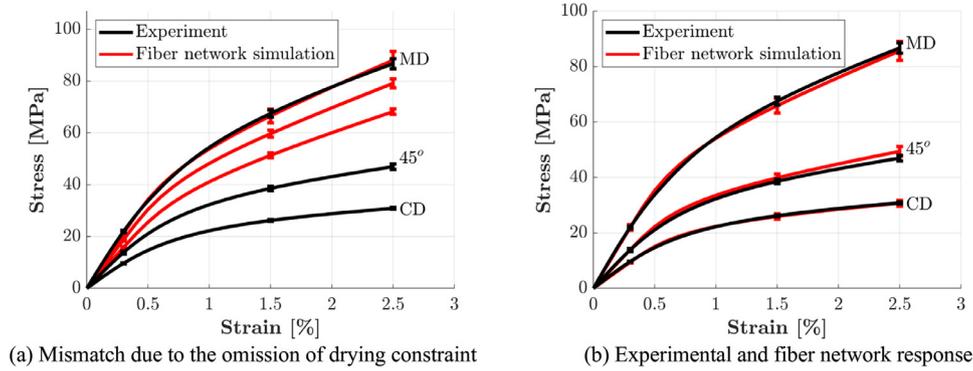

**Fig. 8.** Determination of fiber properties by matching the response of the fiber network simulations to experimental data. Error bars show the standard deviation for different realizations of fiber network simulation and different measurements of the experiment.

In a large paper sample, the anisotropy may vary locally but this variation averages out on the continuum level and does not influence the global stress-strain response of the network as will be demonstrated later. Also, the network may show some small variability owing to its disordered nature even with the same structural parameters. This phenomenon may be size-dependent with smaller samples showing greater variability. To quantify it, we tested 10 networks with the same given set of anisotropy, thickness, and density but different random seeds used during the network generation. We also compared it with the variability observed in the experiment on a larger sample. The average stress-strain curves and the error bars representing one standard deviation are shown in Fig. 8 together with the experimental curves measured on a large sample. These results demonstrate that although the variation exists it is not significant and comparable to the one observed in the experiment. Therefore, the selected fitting parameters are still adequate. It is also important to note, that the greatest variability observed experimentally was in recorded strength values, which are not addressed in this study.

## 4. Mesoscale continuum model

In this Section, we will present a continuum model for the mesoscale response taking into account the fiber anisotropy and density. The fiber orientation and thickness will be accounted for directly in the macro-mechanical finite-element analysis in Section 5.

### 4.1. Plane stress state with hill's yield criterion and isotropic power law hardening

The 2D plane stress model can be written in rate form as $\dot{\boldsymbol{\sigma}} = \mathbf{D}_{tan}\dot{\boldsymbol{\varepsilon}}$ where $\boldsymbol{\sigma} = [\sigma_x, \sigma_y, \tau_{xy}]^T$ is the stress vector, $\boldsymbol{\varepsilon} = [\varepsilon_x, \varepsilon_y, \gamma_{xy}]^T$ is the strain vector and $\mathbf{D}_{tan} = \mathbf{D}_{el} - \mathbf{D}_{pl}$ is the local tangent stiffness matrix computed from the elastic and elastic-plastic stiffness matrix $\mathbf{D}_{el}$ and $\mathbf{D}_{pl}$, respectively. These are defined in a local co-ordinate system where the x-axis points in the fiber orientation direction following the definition in Fig. 3 and Fig. 5(c). The elastic stiffness matrix with a plane stress condition can be written as

$$\mathbf{D}_{el} = \begin{bmatrix} \frac{E_x^2}{E_x - v_{xy}^2 E_y} & \frac{v_{xy} E_x E_y}{E_x - v_{xy}^2 E_y} & 0 \\ \frac{v_{xy} E_x E_y}{E_x - v_{xy}^2 E_y} & \frac{E_x E_y}{E_x - v_{xy}^2 E_y} & 0 \\ 0 & 0 & G_{xy} \end{bmatrix}, \quad (3)$$

where the elastic modulus in x and y direction, $E_x$ and $E_y$, respectively, as well as the Poisson's ratio $v_{xy}$ and shear modulus $G_{xy}$ are functions of the local properties. The Poisson's ratio is given by the empirical expression [52]

$$v_{xy} = 0.293\sqrt{E_x/E_y}, \quad (4)$$

where the constant 0.293 is the Poisson's ratio for the isotropic case. The shear modulus is computed from $E_x$, $E_y$ and $E_{45}$, where the latter is the elastic modulus in the $45°$ direction with respect to the x-axis, according to [53]

$$G_{xy} = \frac{E_x E_y E_{45}}{2v_{xy} E_y E_{45} - E_x E_{45} - E_y E_{45} + 4E_x E_y}. \quad (5)$$

The plastic behavior of the thin fiber network is assumed to follow Hill's yield criterion [54], which is expressed based on the ratio $R_{ij}$ of the yield stress in the direction $ij$ with respect to a reference direction. Choosing the x-axis as the reference direction, i.e. $R_{xx} = 1$, and setting $R_{zz} = 1$, results in the 2D Hills criterion

$$f(\boldsymbol{\sigma}) = \sqrt{\boldsymbol{\sigma}^T \mathbf{H} \boldsymbol{\sigma}} - \sigma_f = 0, \quad (6)$$

where $\sigma_f$ is the flow stress and $\mathbf{H}$ is the Hills orthotropic coefficient matrix given by

$$\mathbf{H} = \begin{bmatrix} 1 & -\frac{1}{2R_{yy}^2} & 0 \\ -\frac{1}{2R_{yy}^2} & \frac{1}{R_{yy}^2} & 0 \\ 0 & 0 & \frac{3}{R_{xy}^2} \end{bmatrix}. \quad (7)$$

The flow stress $\sigma_f$ is given by a power-law hardening

$$\sigma_f = \sigma_s + c\varepsilon_{e,pl}^{1/d}, \quad (8)$$

where $c$ and $d$ are power-hardening constants, $\sigma_s$ is the initial yield stress and $\varepsilon_{e,pl} = \int \sqrt{\frac{2}{3} d\varepsilon_{ij,pl} d\varepsilon_{ij,pl}}$ is the equivalent plastic strain computed from the plastic strain increments $d\varepsilon_{ij,pl}$. It should be noted that Eq. (8) is fitted to the response in the x-direction since $R_{xx} = 1$. During plastic flow, the flow stress $\sigma_f$ is equal to the von Mises effective stress $\sigma_e = \sqrt{\frac{3}{2} s_{ij} s_{ij}}$, where $s_{ij}$ is the deviatoric stress component. An isotropic hardening is assumed, resulting in the following expression for the elastic-plastic stiffness matrix

$$\mathbf{D}_{pl} = \frac{\mathbf{D}_{el} \frac{\partial f}{\partial \boldsymbol{\sigma}} \frac{\partial f}{\partial \boldsymbol{\sigma}}^T \mathbf{D}_{el}}{\frac{\partial \sigma_f}{\partial \varepsilon_{e,pl}} + \frac{\partial f}{\partial \boldsymbol{\sigma}}^T \mathbf{D}_{el} \frac{\partial f}{\partial \boldsymbol{\sigma}}}, \quad (9)$$

where the derivatives of the yield surface and flow stress are computed from Eq. (6) and Eq. (8) as $\frac{\partial f}{\partial \boldsymbol{\sigma}} = \mathbf{H}\boldsymbol{\sigma}/\sqrt{\boldsymbol{\sigma}^T \mathbf{H}\boldsymbol{\sigma}}$ and $\frac{\partial \sigma_f}{\partial \varepsilon_{e,pl}} = \frac{c}{d} \varepsilon_{e,pl}^{1/d-1}$.

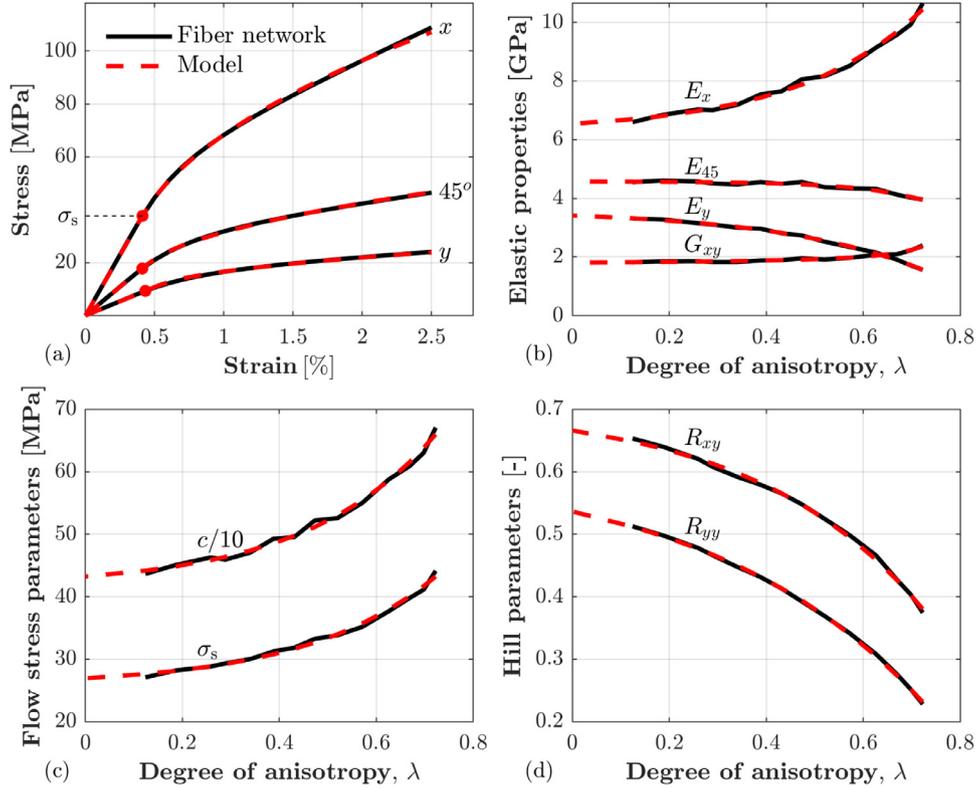

**Fig. 9.** Mesoscale fiber network simulation and proposed continuum model: a) Stress-strain curves for $\lambda = 0.6$, b) elastic properties, c) flow stress parameters and d) Hill's parameters.

### 4.2. Material properties as a function of local anisotropy

The local tangent stiffness matrix derived in Section 4.1 depends on the local anisotropy $\lambda$. This dependency is modeled based on the uniaxial mesoscale fiber network responses in the local $x$, $y$ and 45° directions. In Fig. 9(a), the fiber network responses are shown for $\lambda \approx 0.6$, thickness $t \approx 70$ $\mu$m and density $\rho \approx 1000$ kg/m$^3$ together with the fitted model presented in Section 4.1. In each direction, the elastic modulus is defined as the slope of the line passing through the origin and the point corresponding to a 0.01% plastic strain, where the latter is computed from the initial slope of the curve. The intersection between the stress-strain curve in MD and a line with slope equals to the elastic modulus in MD with 0.01% offset of plastic strain defines the initial yield stress $\sigma_s$. The power-law hardening according to Eq. (8) is thereafter fitted to the plastic part of the response in the $x$-direction, i.e. $\sigma_x \geq \sigma_s$, with respect to the constants $c$ and $d$. The Hill's parameter $R_{yy}$ and $R_{xy}$ are determined by minimizing the least square error of the model response in $y$-direction and 45° direction, respectively. This material determination procedure from the fiber network simulations is repeated for different degrees of anisotropy. The resulting elastic properties ($E_x, E_y, E_{45}$), flow stress parameters ($\sigma_s, c$) and Hill's parameters ($R_{yy}, R_{xy}$) are shown in Fig. 9(b),(c) and (d), respectively, together with the exponential metamodel

$$f(\lambda) = k_1[k_2(\exp(\lambda k_3) - 1) + 1]. \quad (10)$$

The shear modulus computed using Eq. (5) is also shown in Fig. 9(b). In Eq. (10), $k_1$, $k_2$ and $k_3$ are fitting constants such as $k_1$ is the parameter value for $\lambda = 0$, $k_2$ is a dimensionless scale factor and $k_3$ is a dimensionless growth factor. The fitted values are presented in Table 3 for each material parameter. The sign of $k_2$ determines if the material property is increasing ($k_2 > 0$) or decreasing ($k_2 < 0$) with increased degree of anisotropy $\lambda$. The physical meaning of $k_2 < 0$ is that, as the degree of anisotropy increases, less fiber will be in oriented in the CD resulting in a decrease in $E_y$, $R_{yy}$ and $R_{xy}$. From Table 3 it is noted that the hardening exponent $d$ is assumed to be constant, since $k_3 = 0$. Although $d$ is not completely constant, its change with respect to $\lambda$ is small (below 4%) and therefore neglected. It allowed reducing the number of varied parameters while retaining the good quality of the fit across all values of $\lambda$. It is also noted that, during the fitting procedure, the parameters are treated as independent. However, for physical reasons, a certain correlation between the parameters exists. For instance, as the degree of anisotropy increases, more fibers are oriented in MD and fewer fibers in CD. This results in both an increase of $E_x$ and a decrease of $E_y$. Due to the restrained drying condition in MD, the increase in $E_x$ is not equal to the decrease in $E_y$ and the parameter fitting is therefore performed for each individual parameter independently.

### 4.3. Material properties as a function of density

The effect of density $\rho$ is modeled by linear scaling [55] of the elastic moduli $E_x$, $E_y$ and $E_{45}$, the yield stress $\sigma_s$ and the power law hardening constant $c$. The corresponding fitting parameters $k_1$ in Table 3 is multiplied by the density coefficient

$$k_\rho = \frac{\rho - \rho_{per}}{\bar{\rho} - \rho_{per}}, \qquad (11)$$

where $\bar{\rho}$ is the average density of the sheet and $\rho_{per} \approx 200$ kg/m³ [55] is the percolation point of the density. A density lower than $\rho_{per}$ corresponds to insufficient number of fibers for establishing the connectivity across the network.

**Table 3**
Fitting constants $k_1$, $k_2$ and $k_3$ in Eq. (10), determining the influence of anisotropy on the mesoscale material properties.

|  | $E_x$[MPa] | $E_y$[MPa] | $E_{45}$[MPa] | $\sigma_s$[MPa] | $c$[MPa] | $d$ | $R_{yy}$ | $R_{xy}$ |
|---|---|---|---|---|---|---|---|---|
| $k_1$ | 6537 | 3413 | 4570 | 26.93 | 432.2 | 2.023 | 0.5363 | 0.6662 |
| $k_2$ | 0.03998 | −0.05566 | −0.0006364 | 0.04429 | 0.03724 | 0 | −0.1417 | −0.06156 |
| $k_3$ | 3.831 | 3.285 | 7.443 | 3.716 | 3.762 | 0 | 2.231 | 2.872 |

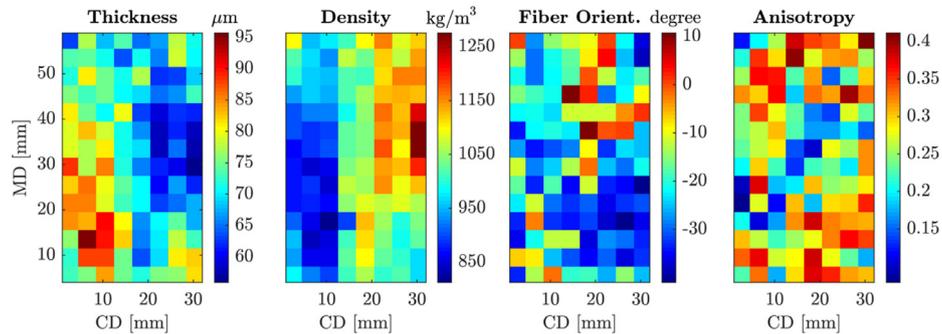

(a) Specimen A

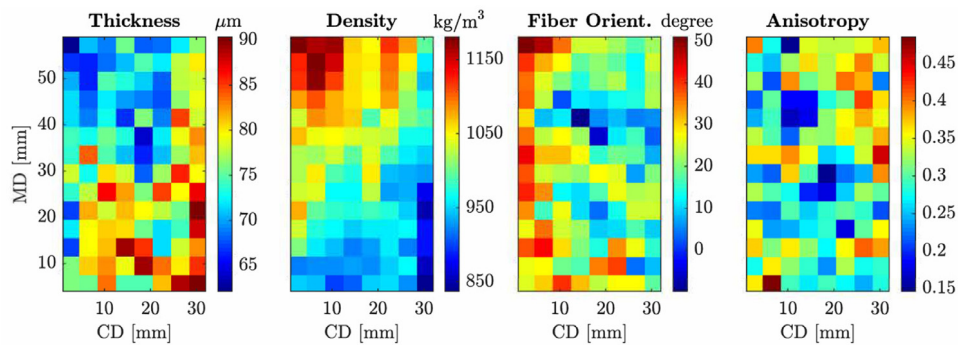

(b) Specimen B

**Fig. 10.** Measured spatial fields after 4 × 4 mm² local averaging. The spatial fields are used as input in the macromechanical FE-model. (For interpretation of the references to colour in this figure legend, the reader is referred to the web version of this article.)

## 5. Model validation and impact of variability

In this section, the spatial strain distribution and the stress-strain response predicted by macro-mechanical FE-simulations are first validated against the experimental results. Thereafter, the impact of local structural variability is evaluated.

### 5.1. Comparison of simulation and experiment

The local variations in the measured structural properties are averaged over the 4 × 4 mm² mesoscale size and inputted in the FE-analysis as can be seen in Fig. 10 for specimen A and B, respectively. It shows that the local averaging introduces discontinuities across the domain. Each 4 × 4 mm² is further meshed with the mesh size of 1 × 1 mm².

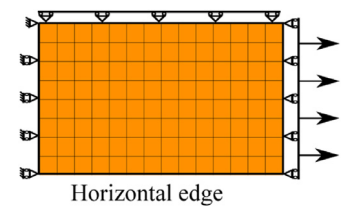

**Fig. 11.** Macromechanical FE-model with applied boundary conditions of a 32 × 56 mm² specimen.

These spatial fields of thickness, density, fiber orientation and fiber anisotropy are used as input in a FE shell model with boundary conditions according to Fig. 11. Each finite-element is governed by the mesoscale continuum model presented in Section 4 with elemental thickness input and local coordinate system pointing in the local fiber orientation direction. The orientation with respect to the drying direction is set to be constant throughout the sheet.

Fig. 12(a) shows the comparison between the 4 stress-strain curves from the characterized samples and the simulated results. Although the curves are relatively close, the simulated results are lower and there is less variability in them. The reason for this is that the examined samples did not have the average Fiber Orientation (FO) equal to zero, as was assumed during the fitting. Since the samples were extracted from arbitrary positions across the width, the average FO was affected by the fact that the forming section of the paper machine may not ensure that the FO is strictly in the MD. It deviates from zero toward the edges of the web. By observing the results, one can see that paper with FO closer to the MD is stiffer and those with the greatest FO are more compliant. The fitting of the micromechanical model was done against an independent set of measurements on the samples with unknown fiber orientation and the assumed FO was zero. With this data fitted, and the coordinate system rotated during computation, this yielded a softer average response compared to the original fitting. To improve the match, one should have accounted for the individual fiber orientation for each sample during the fitting while keeping the drying direction along the MD. In this case, however, one should have the curves in all three directions (MD, CD, and 45°) for a given FO, and getting this data for the same specimen is impossible as the testing in each direction surpasses the yield limit. Nevertheless, this result shows the importance of knowing the fiber orientation for meaningful comparison and its impact on the stress-strain curves, which will be further explored in Section 5.2.

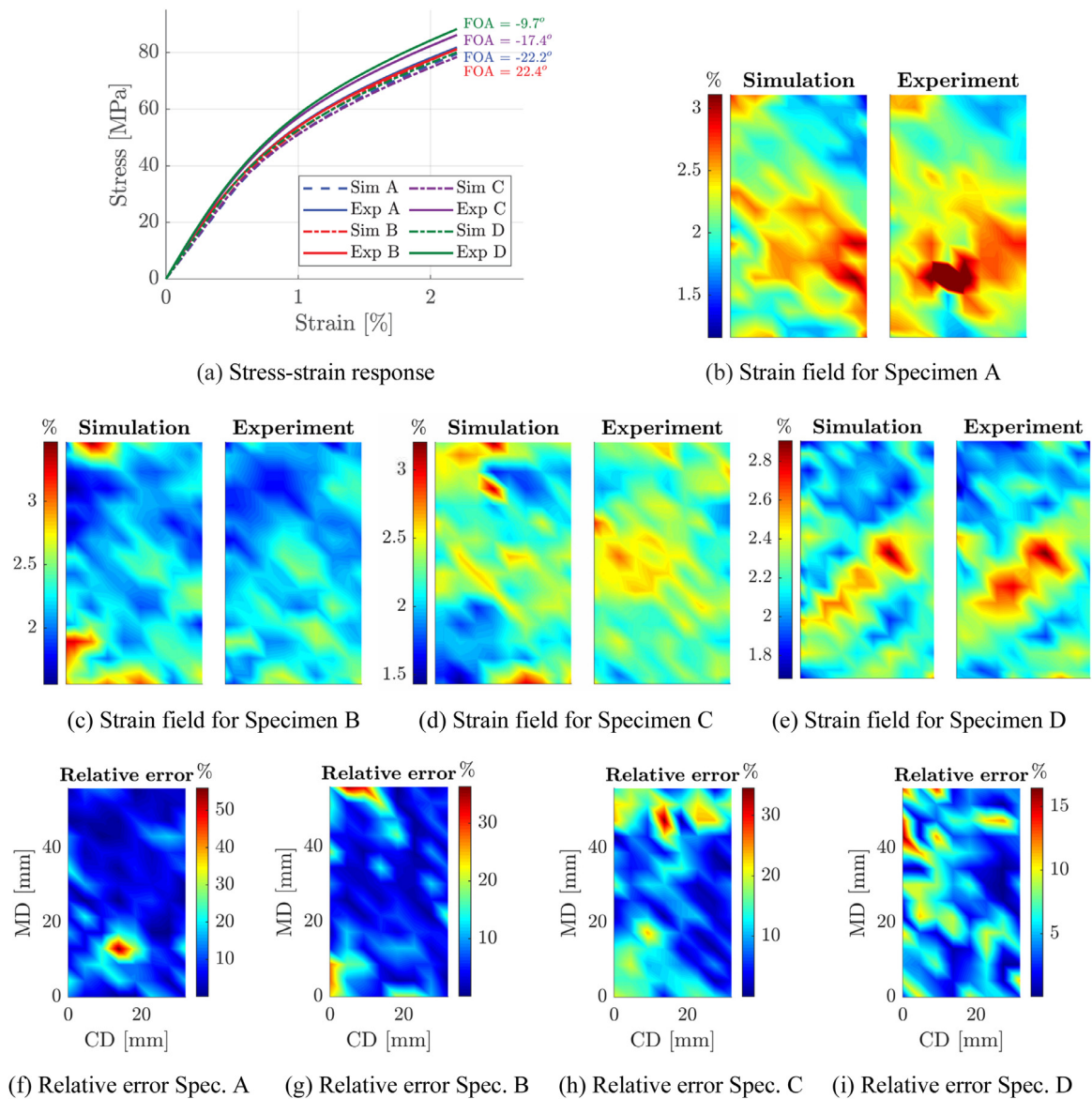

**Fig. 12.** Comparison of experiment and macromechanical FE-simulation a) stress-strain responses, (b-e) measured and simulated spatial strain fields and (f-i) relative errors in simulated strain fields. (For interpretation of the references to colour in this figure legend, the reader is referred to the web version of this article.)

Fig. 12 (b-e) shows the comparison between the computed and measured strain fields. In addition to the visual comparison of the strain fields, relative error plots quantifying the difference between the measured and simulated strain fields are shown in Fig. 12 (f-i). They show an overall good agreement in samples B and D. In samples A and C, the agreement degrades locally where the strain localization takes place, while it remains good outside the localizations. The inability to account for strain softening that may take place locally is a plausible reason for not capturing the localization accurately. The point-wise Pearson correlation coefficient $r$ [56] of measured and simulated strain fields are computed to $r = 0.62$ for Specimen A, $r = 0.60$ for Specimen B, $r = 0.40$ for Specimen C and $r = 0.80$ for Specimen D. Therefore, the multiscale FE-model is capable to capture the regions with higher strains and the regions which are almost intact relatively well.

### 5.2. Influence of local variations

The spatial variability of each of the four studied structural properties is quantified by their Coefficient of Variation (COV) de-fined as the standard deviation of the spatially varying property divided by its mean value. The influence of this variability on the strain field is studied as follows. The spatial variations of thickness, density, fiber orientation and anisotropy are excluded one by one by setting these properties equal to their mean values $\bar{t}$, $\bar{\rho}$, $\bar{\theta}$ and $\bar{\lambda}$, respectively. The corresponding simulated strain fields resulting from uniaxial loading at 2.2% strain are shown in Fig. 13(a) for Specimen A and Fig. 13(b) for Specimen B. The Reference strain fields in Fig. 13 are the simulated strain fields in Fig. 12(b) and (c), which are added for a simpler visual comparison. To quantitatively assess the impact of spatial variability on the strain field, the point-wise Pearson correlation coefficients $r$ between the Reference strain field and each of the other strain fields are computed. The impact of variability, defined as $1 − r$, is plotted in Fig. 14(a) as a function of the experimentally measured COV of each of the four structural properties for the four studied specimens. A linear regression with corresponding 95% confidence interval reveals that the spatial variability in density presents the highest impact on the strain field followed by the thickness and fiber orienta-tion, while the impact of spatial variability in fiber anisotropy is small. The linear regression starts from the origin which represents the theoretical case where the structural property is spatially constant (COV=0) and consequently implying that there is no influence of spatial variability. It is noted that, even though the thickness and density both linearly scale the elemental properties in the FE-implementation, the higher influence of the density is predominantly attributed to the percolation point introduced in Eq. (11). It is also noted that the confidence interval is relatively large due to the stochastic nature of the studied material and the limited number of specimens due to the laborious nature of the experimental study.

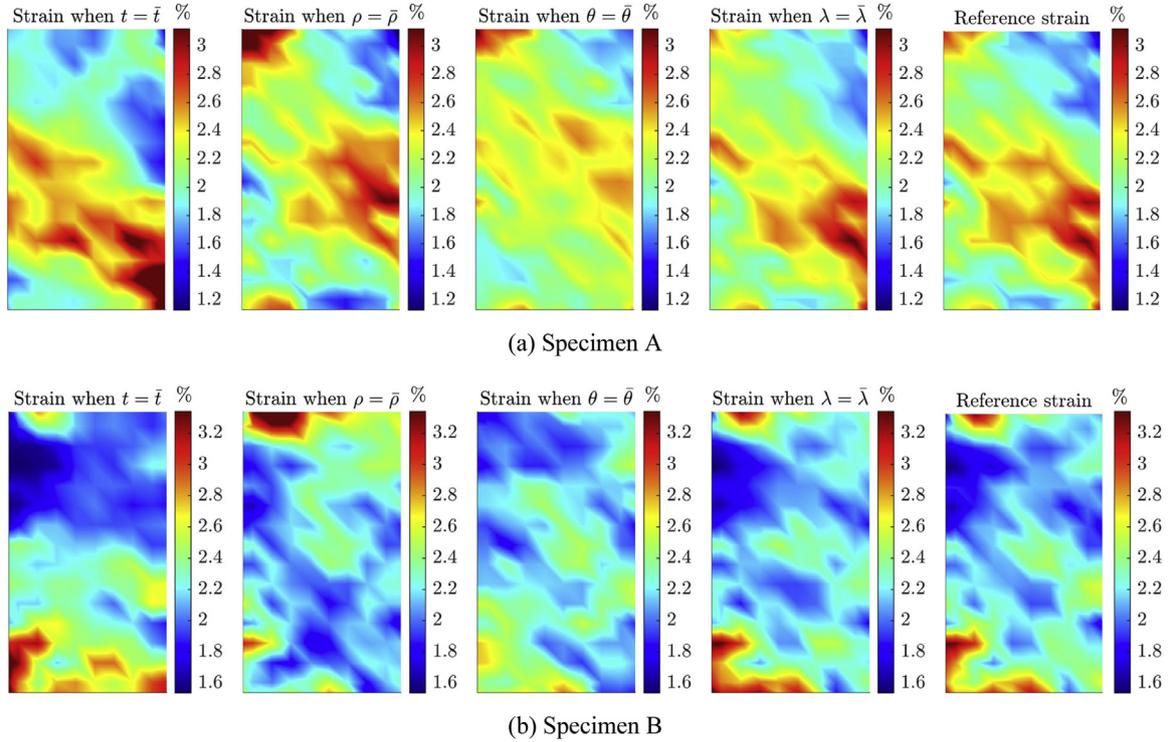

**Fig. 13.** Influence of structural spatial variability on the simulated strain field.

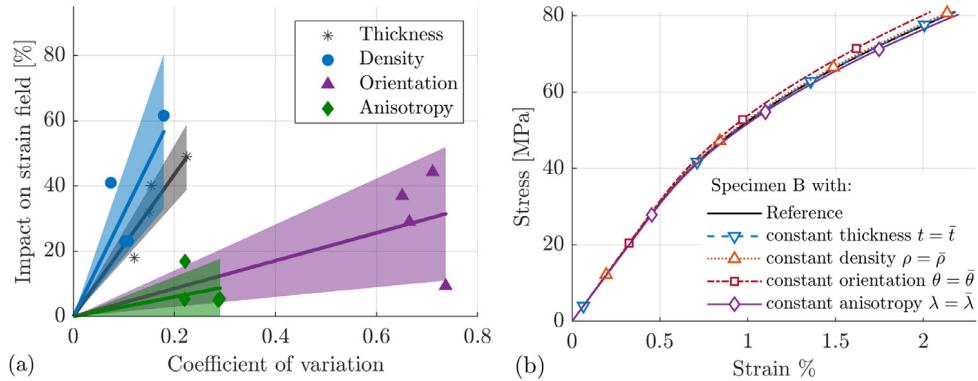

**Fig. 14.** Simulated impact of spatial structural variability a) on the strain field, computed based on 4 specimens with shaded regions corresponding to 95% confidence interval and b) on the macro-mechanical response of Specimen B.

The corresponding impact on the uniaxial response is presented in Fig. 14(b) for Specimen B, where the spatial variations of thickness, density, fiber orientation and anisotropy are excluded one by one by setting these properties equal to their mean values. Similar stress-strain responses are obtained for the other studied specimens. It is noted that the spatial strain fields at 2.2% strain are given in Fig. 13(b). As can be seen, only a small influence of spatial variability in fiber orientation is observed and the impact of variability of the other structural properties is minimal. The results in Fig. 14 also show that fiber orientation variation is somewhat less important to the strain field than to the stress-strain-behaviors. The importance of fiber orientation has already been emphasized in the discussion of the results in Fig. 12(a).

## 6. Discussion

In this work, we developed a combined numerical, modeling and experimental approach to assess the impact of local variability of thin fiber networks. The developed process consists of four major steps.

1. *Micromechanical simulations*. Fiber properties in a micromechanical fiber network model are determined by matching the network response to experimental results. A novelty in this step is the incorporation of the influence of drying conditions during the production process on the fiber properties. Mesoscale fiber network simulations are thereafter used to study the influence of fiber orientation and anisotropy.
2. *Multiscale continuum model*. An anisotropic continuum model combined with an empirical meta-model is derived from the *Micromechanical simulations*. The model is used to establish the link between the mesoscale mechanical response and the structural properties.
3. *Measurement of variability*. Experimental measurements of local structural properties are performed. The variability of each structural property is quantified by the cCoefficient of Variation computed from the measured spatial fields.
4. *Finite-Element model*. The experimentally measured spatial fields are used as input in a macromechanical FE-simulation where each element is governed by the *Multiscale continuum model*. The impact of spatial variability is assessed from the FE-simulations by setting each spatial field to be constant and equal to its average value.

The advantage of the above approach is the ability to decouple the individual impact of local structural variabilities on the mechanical behavior of the material. This isolation of impact is difficult to achieve using a pure experimental method. In addition, the proposed approach bridges the gap between detailed computationally expensive micro-mechanical simulation [20] and the continuum approach [57] which omits the heterogeneity of the materials.

Experimental measurements of the strain field upon loading is performed in order to validate the prediction accuracy of the proposed model. The comparison between the simulation results and the experiments shows an excellent agreement for the mechanical response and a good agreement for the strain field. The deviation between the simulated and measured strain field is explained by 1) the experimental measurement uncertainties, 2) the difference between the continuum finite-element and the fiber network in terms of plasticity and bond behavior and 3) the assumption that the effect of drying on the fiber properties, modeled by Eq. (2), is independent of the degree of anisotropy. Although the latter point is admittedly one of the limitations of the present approach, there are no reports available to put this assumption to the test.

## 7. Conclusions

A multiscale methodology is proposed to quantify the influence of spatial variability of structural properties due to the disordered nature of fiber networks. The proposed method combines detailed micro-mechanical simulations, physical measurement of fiber-level variability and continuum modeling at the mesoscale level. The experimental validation shows an excellent agreement of the macro-mechanical uniaxial response and a good agreement of the predicted strain field.

By using the proposed methodology, we could, for the first time, separate the effect of the fiber and drying anisotropy. We found that the drying anisotropy plays a major role in contributing to the anisotropy of the mechanical properties.

We used the proposed method to identify the role of the local structural variations on the strain variability and mechanical response in the uniaxial tensile test. Among the tested spatial fields, we found that the variations of density followed by thickness and fiber orientation present the largest impact on the strain field, while the fiber anisotropy does not. For the mechanical response, fiber orientation is the principal influencer while other properties are of less importance.

## Declaration of Competing Interest

The authors declare that they have no known competing financial interests or personal relationships that could have appeared to influence the work reported in this paper.

## Acknowledgement

This project is funded by the European Union's Horizon 2020 research and innovation program under the Marie Skłodowska-Curie grant agreement No764713–FibreNet. The financial support of ÅForsk Foundation grant number 18–585 gratefully acknowledged. The financial support of the Federal Ministry of Economy, Family and Youth and the National Foundation for Research, Technology and Development, Austria, is also gratefully acknowledged. We also thank the industrial partners Mondi, Canon Production Printing, Kelheim, and SIG Combibloc for their support. We would like to acknowledge August Brandberg for providing technical support and constructive discussion.

## References

[1] Y. Li, S.E. Stapleton, S. Reese, J.W. Simon, Anisotropic elastic-plastic deformation of paper: out-of-plane model, Int. J. Solids Struct. 130–131 (2018) 172–182.
[2] M. Coelho, L. Hall, J. Berzowska, P. Maes, Pulp-based computing, A framework for building computers out of paper, in: Proc. 27th Int. Conf. Ext. Abstr. Hum. Factors Comput. Syst. - CHI EA '09, ACM Press, New York, New York, USA, 2009, p. 3527.
[3] H. Kröling, A. Endres, N. Nubbo, J. Fleckenstein, A. Miletzky, S. Schabel, Anisotropy of paper and paper based composites and the modelling thereof, in: 16Th Eur. Conf. Compos. Mater., 2014, pp. 1017–1024.
[4] S. Borodulina, H.R. Motamedian, A. Kulachenko, Effect of fiber and bond strength variations on the tensile stiffness and strength of fiber networks, Int. J. Solids Struct. 154 (2018) 19–32.
[5] H. Huang, A. Hagman, M. Nygårds, Quasi static analysis of creasing and folding for three paperboards, Mech. Mater. 69 (2014) 11–34.
[6] D.T. Hristopulos, T. Uesaka, Structural disorder effects on the tensile strength distribution of heterogeneous brittle materials with emphasis on fiber networks, Phys. Rev. B - Condens. Matter Mater. Phys. (2004) 70.
[7] S. Rolland, du Roscoat, M. Decain, X. Thibault, C. Geindreau, J.F. Bloch, Estimation of microstructural properties from synchrotron X-ray microtomography and determination of the REV in paper materials, Acta Mater 55 (2007) 2841–2850.
[8] A. Kulachenko, P. Gradin, T. Uesaka, Basic mechanisms of fluting formation and retention in paper, Mech. Mater. 39 (2007) 643–663.
[9] S.J. I'Anson, W.W. Sampson, S. Savani, Density dependent influence of grammage in tensile properties of handsheets, J. Pulp Pap. Sci. 34 (2008) 182–189.
[10] O. Schultz-Eklund, C.F. P.-Å, Johansson, Method for the local determination of the thickness and density of paper, Nord. Pulp Pap. Res. J. 7 (1992) 133–139.
[11] I. Diddens, B. Murphy, M. Krisch, M. Müller, Anisotropic elastic properties of cellulose measured using inelastic X-ray scattering, Macromolecules 41 (2008) 9755–9759.
[12] H. Hatami-Marbini, R.C. Picu, Effect of fiber orientation on the non-affine deformation of random fiber networks, Acta Mech 205 (2009) 77–84.
[13] C.T. Dodson, W.W. Oba, Yasuhiro and Sampson, On the distributions of mass, thickness and density in paper, Appita J 54 (2001) 385–389.
[14] B. Nordstrom, Effects of grammage on sheet properties in one-sided and two-sided roll forming, Nord. Pulp Pap. Res. J 18 (2003).
[15] R.C. Picu, Mechanics of random fiber networks - A review, Soft Matter 7 (2011) 6768–6785.

M. Alzweighi, R. Mansour, J. Lahti et al.  Acta Materialia 203 (2021) 116460

[16] Y.J. Na, C.L. Muhlstein, Relating nonuniform deformations to fracture in uniaxially loaded non-woven fiber networks, Exp. Mech. 59 (2019) 1127–1144.
[17] M.J. Korteoja, A. Lukkarinen, K. Kaski, D.E. Gunderson, J.L. Dahlke, K.J. Niskanen, Local strain fields in paper, Tappi J 79 (1996) 217–222.
[18] C.T.J.D. Luke Wong, Mark T Kortschot, Effect of formation on local strain fields and fracture of paper., J. Pulp Pap. Sci. (1996) 22.
[19] J.A. Lahti, M. Dauer, D.S. Keller, U. Hirn, Identifying the weak spots in packaging paper: local variations in grammage, fiber orientation and density and the resulting local strain and failure under load, Cellulose (2020).
[20] A. Kulachenko, T. Uesaka, Direct simulations of fiber network deformation and failure, Mech. Mater. 51 (2012) 1–14.
[21] J.W. Simon, A Review of Recent Trends and Challenges in Computational Modeling of Paper and Paperboard at Different Scales, Arch. Comput. Methods Eng. (2020).
[22] R. Bedzra, S. Reese, J.-.W. Simon, Multi-scale modelling of fibre reinforced composites exhibiting elastoplastic deformation, Pamm 17 (2017) 403–404.
[23] B.A. Bednarcyk, B. Stier, J.W. Simon, S. Reese, E.J. Pineda, Meso- and micro-scale modeling of damage in plain weave composites, Compos. Struct. 121 (2015) 258–270.
[24] J. Fish, K. Shek, Multiscale analysis of composite materials and structures, Compos. Sci. Technol. 60 (2000) 2547–2556.
[25] R. Bostanabad, B. Liang, J. Gao, W.K. Liu, J. Cao, D. Zeng, X. Su, H. Xu, Y. Li, W. Chen, Uncertainty quantification in multiscale simulation of woven fiber composites, Comput. Methods Appl. Mech. Eng. 338 (2018) 506–532.
[26] K. Matouš, M.G.D. Geers, V.G. Kouznetsova, A. Gillman, A review of predictive nonlinear theories for multiscale modeling of heterogeneous materials, J. Comput. Phys. 330 (2017) 192–220.
[27] M.F. Hadi, E.A. Sander, V.H. Barocas, Multiscale model predicts tissue-level failure from collagen fiber-level damage, J. Biomech. Eng. (2012) 134.
[28] G. Kettil, A. Målqvist, A. Mark, M. Fredlund, K. Wester, F. Edelvik, Numerical upscaling of discrete network models, BIT Numer. Math. 60 (2020) 67–92.
[29] R. Mansour, A. Kulachenko, W. Chen, M. Olsson, Stochastic constitutive model of isotropic thin fiber networks based on stochastic volume elements, Materials (Basel) 12 (2019) 538.
[30] R. Mansour, M. Olsson, A closed-form second-order reliability method using noncentral chi-squared distributions, J. Mech. Des. Trans. ASME. (2014) 136.
[31] R. Mansour, M. Olsson, Efficient reliability assessment with the conditional probability method, J. Mech. Des. Trans. ASME. (2018) 140.
[32] R. Mansour, M. Olsson, Response surface single loop reliability-based design optimization with higher-order reliability assessment, Struct. Multidiscip. Optim. 54 (2016) 63–79.
[33] Christoph Eberl, Digital Image Correlation and Tracking - File Exchange - MATLAB Central, 2006 https://www.mathworks.com/matlabcentral/fileexchange/12413-digital-image-correlation-and-tracking accessed March 16, 2020.
[34] U. Hirn, M. Lechthaler, W. Bauer, Registration and point wise correlation of local paper properties, Nord. Pulp Pap. Res. J. 23 (2008) 374–381.
[35] D.S. Keller, D.L. Branca, O. Kwon, Characterization of nonwoven structures by spatial partitioning of local thickness and mass density, J. Mater. Sci. 47 (2012) 208–226.
[36] D.S. Keller, J.J. Pawlak, b-Radiographic imaging of paper formation using storage phosphor screens, J. Pulp Pap. Sci. 27 (2001) 117–123.
[37] U. Hirn, W. Bauer, Evaluating an improved method to determine layered fibre orientation by sheet splitting, in: 61st Appita Annu. Conf. Exhib. Gold Coast, 2007, p. 71. Aust. 6-9 May 2007 Proc..
[38] U. Hirn, W. Bauer, A review of image analysis based methods to evaluate fiber properties, Lenzinger Berichte 86 (2006) 96–105.
[39] E.L.G. Wernersson, G. Borgefors, S. Borodulina, A. Kulachenko, Characterisations of fibre networks in paper using micro computed tomography images, Nord. Pulp Pap. Res. J. (2014) 29.
[40] H.R. Motamedian, A.E. Halilovic, A. Kulachenko, Mechanisms of strength and stiffness improvement of paper after PFI refining with a focus on the effect of fines, Cellulose 26 (2019) 4099–4124.
[41] R.C. Neagu, E.K. Gamstedt, F. Berthold, Stiffness contribution of various wood fibers to composite materials, J. Compos. Mater. 40 (2006) 663–699.
[42] T. Pintiaux, D. Viet, V. Vandenbossche, L. Rigal, A. Rouilly, Binderless materials obtained by thermo-compressive processing of lignocellulosic fibers: a Comprehensive review, Bioresources 10 (2015) 1915–1963.
[43] S. Borodulina, A. Kulachenko, S. Galland, M. Nygårds, Stress-strain curve of paper revisited, Nord. Pulp Pap. Res. J. 27 (2012) 318–328.
[44] D. Hart, A.J. Rudman, Least-squares fit of an ellipse to anisotropic polar data: application to azimuthal resistivity surveys in Karst regions, Comput. Geosci. 23 (1997) 189–194.
[45] T. Wahlstrom, Influence of shrinkage and stretch during drying on paper properties, Pap. Technol. 41 (2000) 39–46.
[46] J. Kouko, E. Retulainen, The relationship between shrinkage and elongation of bleached softwood kraft pulp sheets, Nord. Pulp Pap. Res. J. 33 (2018) 522–533.
[47] T. Hansson, C. Fellers, M. Htun, Drying strategies and a new restraint technique to improve cross-directional properties of paper, in: fundam, in: Papermak. Trans. 9 Th Fund. Res. Symp, 1989, pp. 743–781.
[48] P. Page, Derek H. and Tydeman, A new theory of the shrinkage, structure and properties of paper, in: F. Bolam (Ed), Form. Struct. Pap. Trans. Fundam. Res. Symp. Held Oxford, 1961, Tech. sect. Br. Pap. Board MakersÁssoc., 1962.
[49] M.O. Kappil, R.E. Mark, R.W. Perkins, W. Holtzman, Fiber properties in machine-made paper related to recycling and drying tension, ASME Appl. Mech. Div. 209 (1995) 177.
[50] H.L. Cox, The elasticity and strength of paper and other fibrous materials, Br. J. Appl. Phys. 3 (1952) 72–79.
[51] R.P.A. Constantino, S.J. I'Anson, W.W. Sampson, The effect of machine conditions and furnish properties on paper CD shrinkage profile, in: Adv. Pap. Sci. Technol. Trans. XIIIth Fund. Res. Symp, Cambridge, 2005, pp. 283–306.
[52] G.A. Baum, C.C. Habeger, E.H. Fleischman, Measurement of the orthotropic elastic constants of paper, Appleton, Wisconsin : the Institute (1981).
[53] M. Nygårds, Experimental techniques for characterization of elasticplastic material properties in paperboard, Nord. Pulp Pap. Res. J. 23 (2008) 432–437.
[54] R. Hill, A theory of the yielding and plastic flow of anisotropic metals, Proc. R. Soc. London. Ser. A. Math. Phys. Sci. 193 (1948) 281–297.
[55] K. Niskanen, Micromechanics, Mechanics of paper products (Chapt 11), in: K. Niskanen (Ed.), Walter de Gruyter (2011).
[56] T.D.V. Swinscow, M.J. Campbell, Statistics at square one, Bmj London (2002).
[57] D.D. Tjahjanto, O. Girlanda, S. Östlund, Anisotropic viscoelastic-viscoplastic continuum model for high-density cellulose-based materials, J. Mech. Phys. Solids. 84 (2015) 1–20.